%% file: egbib.tex
\documentclass[10pt,twocolumn,letterpaper]{article}

\usepackage{wacv}
\usepackage{times}
\usepackage{epsfig}
\usepackage{graphicx}
\usepackage{amsmath}
\usepackage{amssymb}
\usepackage{hhline}
\usepackage{color, colortbl}
\usepackage{multirow}
\usepackage{times}
\usepackage{cancel}
\usepackage{epsfig}
\usepackage{graphicx}
\usepackage{amsmath}
\usepackage{amssymb}
\usepackage{placeins}
\usepackage{stfloats} 
\usepackage{authblk}
\usepackage{subcaption}
\usepackage[utf8]{inputenc}
\usepackage{lipsum} 
\usepackage{booktabs}
\usepackage{makecell}
\usepackage{url}
\usepackage{algorithm}
\usepackage{algpseudocode}
\usepackage{longtable}
\usepackage{multirow}
\usepackage{cite}
\usepackage{hyperref}
\usepackage{xcolor}
\hypersetup{
    colorlinks,
    linkcolor={red!70!black},
    citecolor={blue!70!black},
    urlcolor={blue!70!black}
}

\def\wacvPaperID{****} 

\wacvalgorithmstrack   

\wacvfinalcopy 

\begin{document}

\title{A Comprehensive Survey on Dynamic Software Updating Techniques in IoTs}

\setlength{\affilsep}{.5em}
\renewcommand\Affilfont{}
\renewcommand\Authsep{\qquad\quad\quad}
\renewcommand\Authands{\qquad}

\author[]{Madhav Neupane \\ Indiana University South Bend \\ {\tt\small \{mneupane\}@iu.edu} }



\maketitle
\thispagestyle{empty}

\begin{abstract}
    This comprehensive survey paper provides an in-depth analysis of Dynamic Software Updating (DSU) techniques in the Internet of Things (IoT). This study critically examines eight significant research papers that employ diverse methodologies to address the challenges of DSU in IoT devices. The primary objectives include comparative analysis to identify the application domains of DSU tools, classification of program alterations accommodated by these systems, evaluation of the advantages and disadvantages of various DSU tools, and identification of potential paths for future research. This paper emphasizes the critical function of DSU in improving energy efficiency, extending operational durability, and bolstering security within IoT environments that demand high availability, including applications in smart cities and connected vehicles. It delves into the basic approaches and mechanisms of DSU, ranging from traditional methods to advanced practices like Over-the-Air updates and container-based solutions. This survey highlights the evolving nature of DSU techniques, balancing operational efficiency, security, and adaptability amidst the complexities of diverse IoT applications. Through this exploration, the paper aims to guide future developments in DSU strategies, enhancing IoT devices' resilience, functionality, and sustainability in a connected world. The insights from this survey are pivotal for researchers, practitioners, and policymakers in shaping effective DSU strategies to meet the growing needs of the IoT ecosystem.
\end{abstract}

\section{Introduction}

The seamless operation of computer systems, particularly in the context of Internet of Things (IoT) devices, is paramount. However, the software maintenance and update landscape poses several challenges, including compatibility issues, system failures, and patch problems. These issues frequently emerge during critical activities such as bug fixes, feature enhancements, configuration modifications, and resource management, which are crucial in IoT devices. In response to these challenges, our research is dedicated to exploring innovative methodologies for software updates that efficiently navigate and mitigate these complexities, ensuring a smooth and effective process for IoT devices. \

Dynamic software updates, often referred to as hot or live updates, encompass modifying or upgrading software applications while they remain actively running without interrupting the operation of IoT devices. This approach eliminates the need to shut down or restart the application, ensuring uninterrupted operation. This capability is crucial for IoT systems with high availability requirements, such as web servers for IoT applications, autonomous vehicles, and critical infrastructure. 
In this comprehensive survey paper, we aim to critically examine eight research papers that employ various tools and methodologies to address the challenges mentioned above in the context of IoT devices. Our primary goals include: 
i) Comparative Analysis: We aim to scrutinize these papers to identify the domains in which dynamic software updating (DSU) tools find application in IoT devices. By doing so, we can gain insights into the diverse contexts where DSU benefits IoT applications. 
ii) Classifying Program Alterations: Our objective is to categorize the various types of program changes that these updating systems can accommodate in the IoT ecosystem. Understanding the range of alterations DSU tools can handle is essential for effectively employing them in different IoT scenarios. 
iii) Evaluating Advantages and Disadvantages: We will assess the strengths and weaknesses of the various DSU tools used in the papers under consideration, focusing on their relevance to IoT devices. This evaluation will provide a comprehensive view of the advantages and limitations of existing DSU solutions in the IoT context. 
iv) Spotting Future Research Paths: Lastly, we will identify areas where future research can contribute to dynamic software updating for IoT devices. This will be based on the patching and updating issues addressed in the papers we examine, suggesting potential directions for further exploration in the rapidly evolving field of IoT technologies. 
Collectively, these studies contribute to the growing field of dynamic software updates in the context of IoT devices, providing valuable insights, tools, and techniques to address the challenges and opportunities in keeping IoT software systems current and secure. As the demand for real-time, uninterrupted operation of IoT devices continues to rise, the research in this domain becomes increasingly relevant and essential.

\section{Importance and Significance of DSU Techniques in IOTs}

Dynamic Software Updating (DSU) techniques in the Internet of Things (IoT) are crucial for sustainable growth and functionality across various sectors. In terms of importance, DSU enhances energy efficiency in IoT devices, particularly in managing Over-the-Air updates, extends operational lifespan in the automotive industry by managing complex software and connectivity, and ensures reliability and safety in real-time embedded systems. DSU is vital for continuous operation and security in high-availability environments like smart cities. The significance of DSU lies in its contribution to the sustainable growth of IoT, enhancing vehicle safety and functionality, maintaining operational continuity in critical systems, facilitating the evolution of smart infrastructure, and providing a dynamic response to new challenges and security threats. DSU's role across these diverse areas underscores its importance in advancing and sustaining IoT technologies.

\section{Basic Approaches and Mechanisms}

This comprehensive survey delves into the various approaches and mechanisms associated with dynamic software updating in the realm of IoT. This encompasses a range of strategies, from fundamental challenges to modern practices, each playing a crucial role in ensuring IoT systems' efficient and secure operation.\\

The paper by Ayres et al. explores various approaches and mechanisms in IoT software updating, addressing both challenges and differences between traditional and modern practices. It highlights significant challenges IoT systems face, akin to those in automotive software, such as code defects and security vulnerabilities, emphasizing the importance of regular updates for system integrity and functionality. While traditional update methods are noted for their reliability, they often lack efficiency and convenience in the context of IoT. On the other hand, modern approaches, particularly Over-the-Air (OTA) updates, provide more streamlined and autonomous processes but come with issues, including security risks and complexities in configuration management. This contrast underlines the necessity for advanced and sophisticated dynamic software updating (DSU) techniques in the IoT landscape \cite{ayres2019virtualisation}.\\

Similarly, Bauwens et al.'s paper discusses that OTA updates in IoT are essential for maintaining and enhancing device functionalities. The process encompasses several steps: update file selection, software module management, and secure software rollout. These steps incur energy consumption, device memory usage, and network traffic overhead, presenting a trade-off between performance gains and update costs  \cite{bauwens2020over}.Furthermore, in Stević et al.'s paper, DSU involves complex procedures to ensure fast, safe, and secure updates in automotive applications. These include evolving software stacks compliant with AUTOSAR standards and incorporating IoT technologies within Adaptive AUTOSAR frameworks. Such updates must consider cloud connectivity, validation, and upgrade procedures for hosted and remote components \cite{stevic2018iot}.\\

In addition, according to the paper by Wahler et al.  \cite{wahler2009dynamic}., Embedded systems in critical industries often forego software updates due to high costs and operational risks. Nevertheless, updates become necessary to adapt to new communication standards, incorporate advanced engineering know-how, and address increasing software complexities and potential bugs. The paper by Mugarza et al. \cite{mugarza2019dynamic}  on smart energy management systems within smart cities highlights that traditional software update methods often necessitate system shutdowns, which are impractical in high-availability applications. Dynamic Software Updating (DSU) techniques overcome this challenge by enabling updates during runtime without compromising system availability and addressing security vulnerabilities and privacy concerns\\

In the paper by Ghosal et al., diverse strategies concerning the security of Over-The-Air (OTA) updates and distributed scheduling have been investigated. The explored approaches encompass Secure Software Repository Frameworks such as Uptane, Blockchain-Based Solutions exemplified by INFUSE, utilization of Cryptographic Hash Functions for update security, the integration of Symmetric and Asymmetric Key Cryptography demonstrated by SecUp, protocols based on Hardware Security Modules (HSM), and the application of Attribute-Based Encryption (ABE) techniques, with STRIDE being a specific example. The comprehensive examination of these methodologies addresses the multifaceted challenges associated with secure OTA updates and distributed scheduling \cite{ghosal2022secure}.\\

The paper by Mugarza et al. on smart energy management systems within smart cities highlights that traditional software update methods often necessitate system shutdowns, which are impractical in high-availability applications. Dynamic Software Updating (DSU) techniques overcome this challenge by enabling updates during runtime without compromising system availability and addressing security vulnerabilities and privacy concerns \cite{mugarza2019dynamic}.\\

The paper by Chi et al. discusses key strategies in dynamic software updating (DSU) for IoT, focusing on dissemination and traffic reduction and the role of the execution environment. Dissemination strategies, such as Directed Diffusion and MOAP, are emphasized for efficiently distributing update packets in environments lacking pre-established infrastructure. Traffic reduction is achieved by transmitting only the modified parts of the software, which helps conserve bandwidth and energy. Regarding the execution environment, DSU typically operates on a hardware abstraction layer provided by operating systems or middleware, enhancing portability across different hardware platforms. Additionally, virtual machines like Maté are highlighted for their benefits in reducing code size and simplifying reprogramming processes \cite{chi2011uflow}.\\

The paper by Kolomvatsos focuses on advanced mechanisms for dynamic software updating in IoT systems, emphasizing three key aspects. First, it advocates for incremental updates, which involve applying small, gradual changes to the software, thus minimizing disruption and resource utilization. Second, it highlights autonomous update mechanisms, enabling IoT devices to independently decide when and how to apply updates, enhancing their self-sufficiency. Lastly, the paper discusses distributed management schemes, which allow individual nodes within the network to manage their update processes, thereby improving the overall resilience and efficiency of the system. Collectively, these mechanisms aim to optimize the software updating process in IoT environments, balancing efficiency with system stability and autonomy \cite{kolomvatsos2018intelligent}.\\

In summary, the survey ``A Comprehensive Survey on Dynamic Software Updating Techniques in IoTs" underscores the diverse and complex nature of DSU in IoT. DSU ensures IoT systems' smooth, secure, and efficient functioning across various sectors, from tackling basic challenges to employing sophisticated mechanisms.

\input{figures_tex/fig1}

\section{Proposed Solutions} \label{PROPOSED SOLUTIONS}

In the context of this survey paper, several proposed and in-use application-level protocols, standards, and solutions are highlighted to address the challenges and requirements of dynamic software updating in IoT systems. These range from container-based updates to specialized operating systems and middleware solutions, each playing a crucial role in advancing the field of IoT software maintenance and enhancement.\\

The paper by Ayres et. al \cite{ayres2019virtualisation} has proposed Container-Based Updates in IoT. Container technology has been proposed for IoT devices to make software updates more accessible, efficient, and robust. This approach addresses the limitations of traditional update mechanisms by offering solutions like rolling updates, live code reloading, and enhanced portability. Containers offer an isolated environment for each application, ensuring that software updates can be managed and deployed more effectively.\\

Furthermore, Bauwens et al. \cite{bauwens2020over} proposed that OTA updates in IoT involve a sophisticated mix of protocols and techniques. These encompass advanced security measures to protect data integrity and confidentiality, efficient dissemination strategies to minimize energy consumption, and reliable installation and activation processes. These methods ensure consistency and stability across the IoT network, addressing the unique challenges of IoT environments. Workflow to perform over-the-air updates, split between the management of software modules and secure software rollout as in given Figure 1.\\

In addition, the paper by Stević et al. \cite{stevic2018iot} proposed the following Automotive IoT Solutions:
a) Adaptive AUTOSAR: Emerging as a standard in automotive IoT, it offers a unified API and addresses the issues of vertically aligned architectures.
b) OTA Update Systems: Proprietary solutions like Ford's Sync 3, Harman's OTA system, and Uconnect are used for automotive software updates.
c) IoT-Based Software Update Proposals: Integration of IoT technologies with Adaptive AUTOSAR and specific installation flows are being considered for next-generation automotive middleware stacks.\\

The paper by Wahler et al. \cite{wahler2009dynamic} proposed:
Operating System Features: OSes like Wind River VxWorks, Green Hills Integrity, and QNX Neutrino support DSU with features like remote debugging and separate address spaces for threads.
 Component Framework: This approach separates components and indirect communication, allowing dynamic replacement of software modules in real-time systems.\\

The paper by Mugarza et al. \cite{mugarza2019dynamic} proposed the Cetratus Framework for Industrial Control Systems. Introduced as a solution for DSU in industrial control systems, the Cetratus framework facilitates live updates without system shutdowns. It aligns with industrial standards and employs an indirection handling table for component management, supporting state transformation for updates.\\

The paper by Chi et al. \cite{chi2011uflow} proposed the following innovative protocols and solutions in IoT:
uFlow: A programming paradigm for Wireless Sensor Networks (WSNs), enabling software updates without rebooting.
Maté Virtual Machine: Designed for sensor networks, focusing on bytecode efficiency.
Neutron: A specialized version of TinyOS that reboots only the faulting units, conserving time and resources.\\

The paper by Kolomvatsos \cite{kolomvatsos2018intelligent} proposed the following protocols and standards for IoT DSU:
Constrained Application Protocol (CoAP): Used for interaction among IoT devices.
Middleware Solutions: Platform-independent middleware that manages heterogeneity in IoT devices, ensuring data transformation and connectivity.\\

These protocols, standards, and solutions represent diverse approaches to addressing IoT's dynamic software updating needs. They exemplify the evolving landscape of IoT software maintenance, highlighting innovations and emerging trends tailored to the unique challenges of IoT systems.\\

\input{figures_tex/fig2}

\section{Classification of Approaches} \label{CLASSIFICATION OF APPROACHES}
Based on the contents of the paper by Ayres et al. \cite{ayres2019virtualisation}, the current approaches to dynamic software updating techniques in IoTs, particularly in the automotive sector, can be classified as follows:
Traditional Update Methods: Encompass manual, scheduled downtime, and OTA (Over-the-Air) updates are widely used in current IoT environments, especially for routine tasks.
 Advanced IoT Update Methods: Utilize VM-based and container-based updates for secure, efficient, and isolated software updating in complex systems like automotive E/E architectures.\\

In the paper by Bauwens et al. \cite{bauwens2020over}, OTA software updates can be classified into two main categories:
Full Firmware Updates: These updates involve replacing the entire code base of the IoT device. While comprehensive, they require substantial bandwidth and energy, resulting in higher latency.
Modular Updates: These are partial, targeting specific components like the MAC layer. They are more energy-efficient compared to full firmware updates.\\

The paper by Stević et al. \cite{stevic2018iot} can be classified as:
 Traditional Update Managers
            Systems like OSTree and GENIVI SOTA represent traditional update managers, which are not always applicable to life-critical automotive ECUs.
 IoT-Integrated Update Systems
            New approaches integrate IoT technologies for OTA updates, as seen in the OBLO
            system and Adaptive AUTOSAR platform, enhancing the update process and 
            addressing safety and security concerns. Proposed architecture of the Software Update is shown in Figure 2. \\

In the paper by Wahler et al. \cite{wahler2009dynamic}, the approaches can be classified as follows:
a) Routine-Based Updates
These are fine-grained updates for individual routines, requiring runtime support and code analysis for type safety.
b) Component-Based Updates
These updates are more coarse-grained, allowing for the replacement of components at runtime. This approach ensures isolation between components and allows for online reconfiguration.\\

The paper by Mugarza et al. \cite{mugarza2019dynamic} proposes the following approaches:
a) Hardware Redundancy-Based Updates
    Involves secondary hardware platforms for loading new software versions, allowing for role changes without disrupting operations. It is typically employed in systems where continuity and high availability are critical.
b) Software-Driven Dynamic Updates
    Direct modifications to running software are made without requiring hardware redundancy or system restarts, transforming the currently running program to a new version through code and state transformations. It is more prevalent in IoT and IIoT systems, especially smart city applications.\\

In the paper by Ghosal et al. \cite{ghosal2022secure}, the classification of current approaches to secure Over-the-Air (OTA) update techniques in IoT, particularly for connected vehicles, focuses on several key aspects:
Scalable Scheduling: This feature, present in many studies, allows for effective management of update schedules across numerous devices, which is essential for more extensive IoT networks. However, some works need to incorporate this, potentially limiting their efficiency in complex systems.
Authentication: Most approaches prioritize authenticating the source of software updates, a crucial factor for ensuring the security and reliability of the update process. One study notably needs to emphasize this aspect.
Confidentiality: Many techniques ensure confidentiality to protect the data involved in the update process from unauthorized access, with a few exceptions that do not focus on this.
Integrity: Nearly all the referenced works maintain the integrity of software updates, ensuring the data remains unchanged and secure during transmission.
Data Freshness: Only the proposed technique in the paper uniquely emphasizes data freshness, ensuring that recent updates are safeguarded against outdated or replayed versions, a feature not commonly addressed in other studies.\\

Based on the paper by Chi et al. \cite{chi2011uflow}, the current approaches to software updating in wireless sensor networks can be classified into three key perspectives:
Dissemination-Based Approaches:
           This perspective focuses on effectively spreading the software update message across
           the network. It is essential in wireless sensor networks where nodes often operate
          distributed without pre-established infrastructure for routing and forwarding information.
Traffic Reduction Strategies:
            This approach deals with creating the software patches necessary for updates. It is
            centered around generating and distributing only the essential parts of the software that 
            need modification, thereby optimizing the update process in terms of bandwidth and 
            storage.
Execution Environment-Based Approaches:
           The execution environment perspective involves the actual implementation and 
           deployment of the software update. It ensures that the updated software operates correctly
           within the sensor network's existing hardware and software infrastructure.\\

The paper by Kolomvatsos \cite{kolomvatsos2018intelligent} introduces a new approach to software updating in wireless sensor networks (WSNs), specifically designed for the dynamic and decentralized nature of IoT environments. It challenges the traditional centralized methods that rely on a central server for disseminating updates, which often face issues with incremental updates, compression, diffusion, and managing broadcast messages. These centralized systems also struggle with node heterogeneity and effective dissemination without overwhelming network bandwidth. The paper's proposed solution is a distributed scheme that leverages the autonomous nature of individual nodes. This model shifts the focus from the method of applying updates to the timing of initiation. Key features of the proposed approach include:
Performance-Aware Mechanism: It considers the load of individual nodes and the overall network performance, initiating updates only when the network is in an optimal state to support efficient and secure updates.
Load Balancing Scheme: This aspect aims to prevent network bottlenecks when updates are available, ensuring a smooth and uninterrupted application process.
Ensemble Forecasting Scheme: The model uses multiple performance metrics to forecast future network conditions, determining the most appropriate time for initiating updates.
Decision-Making Mechanism: Based on current and predicted network and node performance, initiating updates is made using an Artificial Neural Network (ANN), which determines the optimal timing for each node. \\

\input{figures_tex/fig3}

Representation shown in Figure \ref{fig3} provides a hierarchical view of the different categories and subcategories of dynamic software updating techniques, making it easier to understand the relationships between them based on the content of the referenced papers.\\

\section{ Relative Advantages and Disadvantages between the Classified Techniques and Standards}
The paper by Ayres et al. \cite{ayres2019virtualisation} examines various techniques and standards for software updates in connected and autonomous vehicles, focusing on Over-the-Air (OTA) updates, virtualization, and container-based environments. OTA updates significan tly reduce vehicle downtime by avoiding physical interventions, while virtualization improves hardware resource efficiency and takes advantage of multi-core processors. Container-based architectures offer operational efficiency and scalability for continuous updates. However, OTA updates can pose security risks and are not ideal for urgent updates. Due to full operating system requirements, virtualization may lead to cumbersome and vulnerable systems. Container-based systems, though efficient, require meticulous implementation to maintain robustness and reliability, which is crucial for safety and performance in automotive systems.\\

The paper by Bauwens et al. \cite{bauwens2020over} provides an in-depth analysis of the pros and cons of various software updating techniques in IoT environments. Advantages include the pivotal role of Over-the-Air (OTA) updates in IoT evolution for remote modifications, an in-depth and energy-conscious software development and deployment process, comprehensive measures for secure and reliable software update steps, and a secure software rollout phase that ensures data transfer security and coordinated updates. However, these methods have drawbacks: OTA updates significantly increase energy consumption, the software development process is complex and affects operational procedures, software update steps are resource-intensive, pre-deployment verification adds complexity and resource demands and the secure software rollout phase imposes a considerable overhead on device memory, network traffic, and power consumption, which is challenging for battery-powered devices.\\

The paper by Stević et al. \cite{stevic2018iot} explores various techniques and standards for software updates in modern vehicles, highlighting their advantages and disadvantages. The benefits include the convenience of Over-the-Air (OTA) updates for quick software delivery, the dynamic software upgrade capabilities of Adaptive AUTOSAR, and the tailored automotive update solutions of proprietary systems like Ford's Sync 3. Integrating IoT technologies, along with systems like OBLO and ARA::UCM, enhances communication and security in software management. However, these techniques also have drawbacks: OTA updates demand sophisticated management platforms for safe deployment, the development of secure upgrade routines in systems like Adaptive AUTOSAR is an ongoing challenge, proprietary solutions can lead to inflexible software architectures that hinder system upgrades, and the integration of IoT and systems like OBLO and ARA::UCM introduces complexity, necessitating careful management and specific conditions for effective updates.\\

The paper by Wahler et al. \cite{wahler2009dynamic} delves into various techniques and standards for dynamic software updating in real-time systems, particularly in the power and automation sectors. The advantages of these techniques include dynamic software updating in embedded systems to adapt to evolving protocols and requirements, thus enhancing flexibility. Modern real-time operating systems like Wind River VxWorks indirectly support dynamic updates. The Component Framework aids in conducting safe operational updates through component separation, and Real-Time Updates and Algorithms, coupled with State Transfer, minimize performance impact while allowing immediate update applicability. On the downside, these updates face challenges in real-time systems due to strict deadlines and limited resources. Real-time operating systems often lack inherent support for these updates, potentially leading to performance issues. The Component Framework introduces design complexity and restricts direct component interaction. Furthermore, the update process and state transfer in real-time environments are complex and demand careful management to adhere to tight constraints.\\

The paper by Mugarza et al. \cite{mugarza2019dynamic} examines dynamic software updating (DSU) techniques in Internet of Things (IoT) and Industrial Internet of Things (IIoT) systems, analyzing their benefits and drawbacks. DSU techniques are essential for uninterrupted operations in high-availability environments like smart cities, enabling updates without system shutdowns. Hardware Redundancy for Live System Updates ensures smooth transitions using a secondary platform, maintaining continuous primary system operations. The Cetratus Runtime Framework offers a mixed-criticality architecture for dynamic updates, complying with industrial standards and enhancing privacy through advanced encryption. Code and State Transformation Methods provide flexible update processes with innovative code and state modification approaches. However, these techniques also face challenges: DSU implementation is complex, particularly in integrating existing systems and managing component criticality levels. Hardware redundancy is resource-intensive and can lead to inefficiencies. The Cetratus Framework's integration into existing systems can be challenging for security and performance maintenance. Lastly, the complexity of Code and State Transformation Methods requires meticulous management due to significant memory demands and program data size growth \cite{mugarza2019dynamic}. \\

The paper by Ghosal et al. \cite{ghosal2022secure} examines the advantages and disadvantages of various techniques and standards for Over-the-Air (OTA) software updates in connected vehicles. OTA updates are pivotal for these vehicles, offering remote upgrades that are convenient, fast, and cost-effective. The STRIDE technique enhances OTA updates by ensuring secure and scalable software distribution, utilizing cloud technology for efficient delivery while reducing overheads. Other methods, like the Uptane framework and Blockchain-based solutions, improve security, focusing on authenticity, integrity, and confidentiality. Despite these benefits, OTA updates face significant security challenges, including consistent, reliable connectivity issues and varied network behaviors. While offering end-to-end security, STRIDE's implementation is complex due to robust encryption and dynamic scheduling requirements. Uptane, although effective, may be vulnerable to rollback attacks. Blockchain solutions, known for their security, face challenges in source reliability detection and can be resource-intensive, leading to slower update delivery. Furthermore, achieving comprehensive encryption is complex due to the multiple intermediaries between the original equipment manufacturer (OEM) and the vehicle.\\

The paper by Chi et al. \cite{chi2011uflow} explores various software updating approaches in wireless sensor networks, highlighting their pros and cons. Advantages include efficient dissemination techniques like Directed Diffusion and MOAP, effectively spreading update messages in infrastructure-less environments. Like the UNIX diff system, Patch Generating creates small, cost-effective patches, transmitting only essential changes. Execution Environments in software updates facilitate the process by allowing dynamic replacement and task updates without disrupting ongoing operations, aiding in task management and resolving concurrency issues. On the other hand, dissemination methods may require dividing large software images into smaller segments for transmission. This process can be resource-heavy and inefficient for extensive updates. The complexity of patch generation and application is heightened in environments with limited resources, posing challenges in applying patches consistently across different software versions. Execution Environments, though they simplify updates, require skilled management and can lead to additional overheads, particularly when employing resource-demanding virtual machines like Maté.\\

The paper by Kolomvatsos \cite{kolomvatsos2018intelligent} discusses a distributed approach for software updates in the Internet of Things (IoT), comparing its advantages and disadvantages to centralized update management systems. The distributed scheme enhances IoT node autonomy, allowing individual nodes to determine the optimal timing for updates. It incorporates a performance-aware mechanism that accounts for node loads and overall network efficiency, employs load-balancing to avoid bottlenecks, and uses an ensemble forecasting model supported by an Artificial Neural Network (ANN) for informed decision-making. However, this approach adds complexity as each node must independently manage updates, leading to additional computational overhead and potential inconsistencies in update applications due to varying perceptions of network conditions.\\

In contrast, centralized software update management systems face challenges in handling node heterogeneity and network congestion, as central servers bear the heavy load of distributing updates. These systems often struggle with the complexity and performance issues associated with managing a diverse network, emphasizing the limitations of centralized approaches in such environments. The distributed update management scheme offers improved performance optimization and node autonomy. However, it comes with management complexities and potential inconsistencies, contrasting with the centralized system's struggles with network congestion and node heterogeneity.

\section{Some Leading Research Papers}
The paper by Ayres et al. \cite{ayres2019virtualisation} identifies several leading research works, products, and websites that contribute significantly to dynamic software updating techniques in IoTs, particularly in the context of automotive systems. Here are some of the notable references from the paper:
Backhausen et al. \cite{backhausen2017robustness}: "Robustness in automotive electronics: An industrial overview of major concerns," presented at the On-Line Testing and Robust System Design (IOLTS) conference in 2017.
Bello et al. \cite{bello2018recent}: "Recent Advances and Trends in On-board Embedded and Networked Automotive Systems," published in IEEE Transactions on Industrial Informatics, 2018.\\

The paper by Bauwens et al. \cite{bauwens2020over} references several leading research works, products, and websites relevant to dynamic software updating techniques in IoTs. Here are some of the key references from the paper:
H. Guissouma et al. \cite{guissouma2018empirica}, "An Empirical Study on the Current and Future Challenges of Automotive Software Release and Configuration Management," presented at the Euromicro Conference on Software Engineering and Advanced Applications by IEEE in 2018.
P. Ruckebusch et al.\cite{ruckebusch2018modelling}, "Modelling the Energy Consumption for Over–the–Air Software Updates In LPWAN Networks: Sigfox, LoRa, and IEEE 802.15.4g," published in the Internet of Things Journal, vol. 3, in 2018.
J. C. Cano et al. \cite{cano2018evolution}, "Evolution of IoT: An Industry Perspective," published in the IEEE Internet of Things Magazine, vol. 1, no. 2, in 2018.\\

\section{ Unsolved Problems and Future Research Issues}
The paper by Ayres et al. delves into the unresolved issues and challenges associated with dynamic software updating in the Internet of Things (IoTs), focusing on automotive applications. It underscores the complexity of automotive E/E architecture and the difficulties in implementing dynamic updates due to system intricacy. The robustness and reliability of container-based systems are highlighted, particularly the challenges in achieving operational robustness and low latency. The paper also stresses the security vulnerabilities in modern, connected, and autonomous vehicles, emphasizing the need for secure updates. It discusses integrating virtualization technologies, like containers and unikernels, into automotive software as a complex but crucial task. Future research directions include enhancing container-based E/E architecture, exploring lightweight virtualization solutions, developing more efficient software update mechanisms, and improving scalability and resource management in container-based systems. The paper concludes that while container-based virtualization and other emerging technologies show promise for automotive software updates, significant complexity, security, and integration challenges still need to be addressed \cite{ayres2019virtualisation}.\\

The paper by Bauwens et al. tackles the critical challenges of dynamic software updating in the Internet of Things (IoTs), particularly emphasizing the importance of verifying third-party code trustworthiness and handling complexities in multi-owner environments. It underscores the necessity of advanced code isolation techniques to safeguard against malicious code in vulnerable IoT devices. The paper also addresses the challenges of upgrading network stacks, including the physical layers, in the context of advancements in software-defined radio and field-programmable gate arrays. It proposes future research directions that focus on applying software-defined networking and virtualization for managing lower-layer protocol behaviors and the potential of introducing complete software components at the application layer. Additionally, the paper highlights the benefits of edge or fog-based architectures for efficient data dissemination and bandwidth and latency management in large-scale IoT networks, identifying these as key areas for future exploration and development \cite{bauwens2020over}.\\

The paper by Stević et al. explores unresolved challenges and future research directions in dynamic software updating techniques in IoTs, with a specific focus on automotive applications. It underscores the complexity of over-the-air (OTA) updates in automotive systems, particularly the challenges in safely and efficiently integrating these updates within platforms like Adaptive AUTOSAR. The paper emphasizes major concerns around security and safety, especially in protecting Electronic Control Units (ECUs) from external intrusions and ensuring that OTA updates do not compromise the system's safety. For future research, it suggests the development of reliable and safe software upgrade routines within the Adaptive AUTOSAR framework, which includes cloud connectivity integration and validation procedures. The paper also highlights the need to carefully integrate IoT technologies for smart home applications into automotive platforms, addressing cloud services and gateway communication challenges. Optimizing automotive update managers for safety validation, authorization, and data correctness is another crucial area, along with developing rollback procedures for failed updates. Lastly, the paper emphasizes the importance of thoroughly evaluating and testing proposed solutions in simulated environments to ensure their effectiveness and reliability \cite{stevic2018iot}.\\

The paper by Wahler et al. addresses key unresolved issues and future research directions in dynamic software updating, focusing on maintaining the ACID (Atomicity, Consistency, Isolation, Durability) principles during the simultaneous replacement of multiple components. This aspect is crucial for ensuring the transactional integrity of updates. The paper also highlights the challenge of dealing with potentially malicious components that resist resource release and termination, suggesting the need to enhance the framework to revoke resources from such non-compliant components forcibly. Furthermore, it identifies the necessity to optimize the current component framework to reduce the number of context switches during updates, which currently caps system performance at a maximum frequency of 300 Hz. Addressing this limitation through future optimizations is seen as vital for significantly enhancing system performance, making it an essential area for further research and development \cite{wahler2009dynamic}.\\

The paper by Mugarza et al. tackles key unresolved issues and future research areas in IoT and IIoT, focusing on security, privacy, and system architecture. It highlights the growing concerns of security and privacy in interconnected environments and the need for regular software updates, which can be challenging due to the requirement of system shutdowns in high-availability applications like smart city services. The paper proposes a mixed-criticality software architecture for intelligent energy applications, ensuring isolation among its critical components to maintain system integrity and reliability. It introduces the Cetratus framework, enabling dynamic software updates during runtime without necessitating system shutdowns, thus addressing the challenge of continuous service in high-availability systems. Additionally, the paper discusses enhancing security and privacy in smart city applications through an added layer of security using a homomorphic encryption algorithm in a data collector agent. It suggests implementing an access control scheme for robust authentication, authorization, and auditing. It considers using a hypervisor-based approach as an alternative to partitioning-enabled operating systems for further development in this field \cite{mugarza2019dynamic}.\\

The paper by Ghosal et al. identifies two critical unresolved problems in dynamic software updating in IoT environments, particularly for connected vehicles. Firstly, it emphasizes developing a vehicle-specific key revocation mechanism to enhance security against sophisticated attacks that exploit compromised vehicles. Such a mechanism is vital for mitigating threats posed by compromised vehicles. Secondly, the paper points to the importance of investigating how Original Equipment Manufacturer (OEM) failures might impact the performance of the STRIDE technique, especially in high-density connected vehicle environments. Understanding the effects of OEM failure is crucial for ensuring the resilience and reliability of the system under various operational conditions. Addressing these future research directions is key to tackling potential vulnerabilities, thereby bolstering the security and efficiency of dynamic software updating techniques in the context of connected vehicles \cite{ghosal2022secure}.\\

The paper by Kolomvatsos addresses significant challenges in the IoT sector, concentrating on issues like autonomy, load balancing, network interference, and energy consumption. It highlights the difficulty of enabling IoT nodes to independently determine the most suitable timing for software updates while considering network performance. The paper also delves into the challenge of avoiding central server overload by balancing loads and minimizing network interference, which is crucial for maintaining network efficiency and energy efficiency in individual IoT nodes and the broader network. For future research, Kolomvatsos proposes developing an adaptive model designed explicitly for IoT nodes' varying demands and workloads. This model aims to navigate the uncertainties in the nodes' operating environments adeptly. Additionally, the paper underscores the importance of creating a comprehensive model capable of managing updates under diverse environmental conditions and network states to ensure efficient and effective updates across various IoT scenarios. Considering immediate and long-term needs, this approach aims to enhance IoT systems' overall functionality and sustainability \cite{kolomvatsos2018intelligent}.\\

\section{Comparative Analysis}
Table \ref{table1} shown below demonstrates the comparison of the eight papers based on the various categorization. The categorization are listed in the table.

\input{tables_tex/table1}
\input{tables_tex/table2}

\section{Conclusion}
In concluding this comprehensive survey on Dynamic Software Updating Techniques in IoTs, we have explored a wide range of research papers, each contributing distinct perspectives and solutions to the challenges in dynamic software updating for IoT devices. The survey spans various sectors within the IoT domain, including automotive systems, smart cities, real-time embedded systems, and industrial IoT, focusing on addressing these diverse applications' unique challenges and requirements.\\

We scrutinized formal publications across several respected journals and conference proceedings to prepare this survey. Key sources include IEEE Transactions on Industrial Informatics, Internet of Things Journal, and various IEEE conferences such as the Euromicro Conference on Software Engineering and Advanced Applications and the On-Line Testing and Robust System Design (IOLTS) conference. The papers reviewed span a timeframe primarily from the mid-2010s to the present, reflecting the latest trends and advancements in the field.\\

This survey reveals the evolving nature of dynamic software updating techniques in IoT, highlighting the balance between operational efficiency, security, and adaptability. The complexities inherent in the IoT landscape, from energy constraints to diverse operational environments, necessitate continual innovation and research in DSU techniques. Future research paths identified in this survey underscore the need for more secure, efficient, and adaptable updating mechanisms tailored to the specific requirements of various IoT applications.\\

The insights garnered from this survey aim to guide researchers, practitioners, and policymakers in developing and implementing effective DSU strategies that cater to the burgeoning needs of the IoT ecosystem. The ultimate goal is to enhance IoT devices' resilience, functionality, and sustainability, ensuring they remain effective tools in our increasingly connected world.\\


\clearpage
{\small
\bibliographystyle{unsrt}
\bibliography{egbib}
}









\end{document}

%% file: figures_tex/fig1.tex
\begin{figure*}[!ht]
  \centering
  \includegraphics[width=0.75\textwidth,height=5cm]{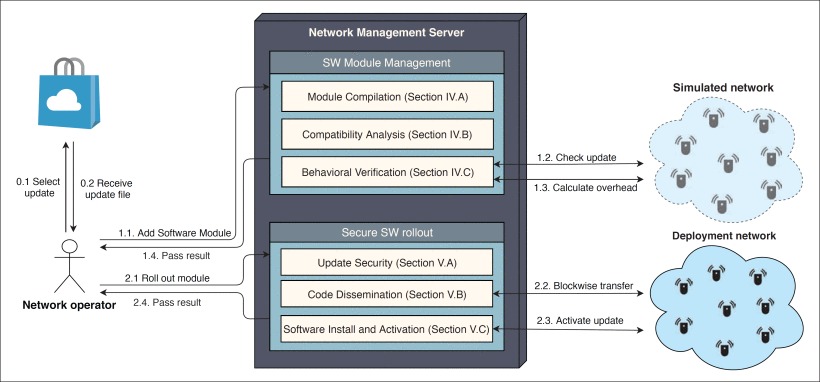}
  \caption{OTA Updates \cite{bauwens2020over}}
  \label{fig:example}
\end{figure*}

%% file: figures_tex/fig2.tex
\begin{figure*}[!ht]
  \centering
  \includegraphics[width=0.75\textwidth,height=7.5cm]{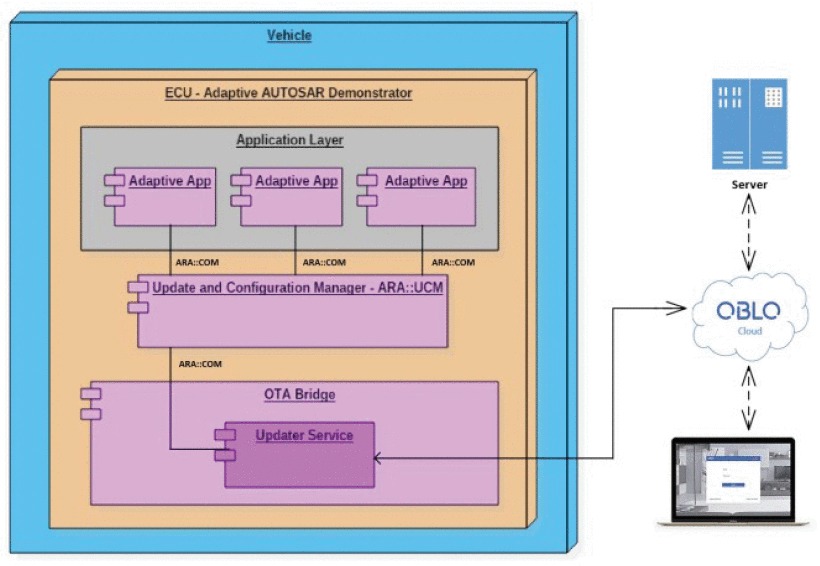}
  \caption{Architecture of Update Flow}
  \label{fig2}
\end{figure*}

%% file: figures_tex/fig3.tex
\begin{figure*}[!ht]
  \centering
  \includegraphics[width=0.95\textwidth,height=8.25cm]{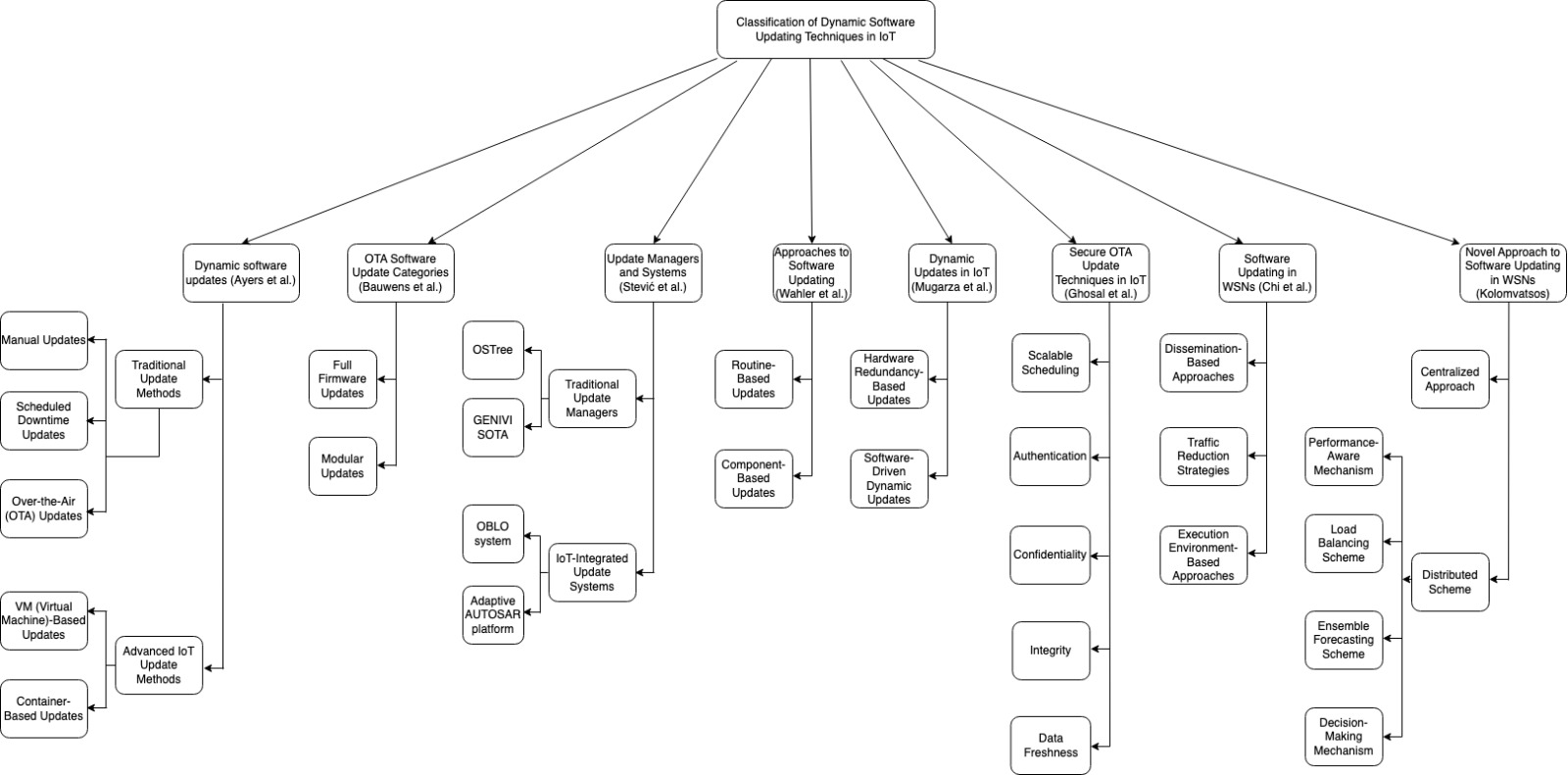}
  \caption{Tree Diagram for Classification of DSU Techniques}
  \label{fig3}
\end{figure*}

%% file: tables_tex/table1.tex
\setlength\extrarowheight{10pt}
\begin{table*}[ht!]
\centering
\resizebox{.95\textwidth}{!}{%
\begin{tabular}{|l|l|l|l|l|}
\hline
{\bf Comparison Matrix $\downarrow$} &
  { \bf  Details from  Ayres et al. Paper $\downarrow$}   &
 {\bf Details from Bauwens et al. Paper $\downarrow$}  &
  {\bf Details from Stević et al. paper $\downarrow$} &
 {\bf  Details from Wahler et al. Paper $\downarrow$ } \\ \hline
Application Domain &
  \begin{tabular}[c]{@{}l@{}}Focuses on automotive systems\\ , particularly connected \\ and autonomous vehicles.\end{tabular} &
  \begin{tabular}[c]{@{}l@{}}Focuses on Internet of Things (IoT)\\  environments, particularly \\ addressing the challenges \\ and methodologies of \\ Over-the-Air (OTA) \\ software updates.\end{tabular} &
  \begin{tabular}[c]{@{}l@{}}Centered on automotive \\ applications, particularly\\  focusing on software updates \\ in modern vehicle systems.\end{tabular} &
  \begin{tabular}[c]{@{}l@{}}Focuses on real-time systems,\\  particularly in power and \\ automation industries.\end{tabular} \\ \hline
Update Methodology &
  \begin{tabular}[c]{@{}l@{}}Traditional update \\ methods and modern approaches \\ like Over-the-Air (OTA) updates, \\ virtualization, and container-based \\ environments.\end{tabular} &
  \begin{tabular}[c]{@{}l@{}}Highlights OTA software updates, \\ detailing their process, from update \\ file selection to secure software rollout, \\ including the management of \\ software modules.\end{tabular} &
  \begin{tabular}[c]{@{}l@{}}Examines Over-the-Air (OTA) updates \\ and the integration of IoT \\ technologies within the Adaptive \\ AUTOSAR framework, highlighting \\ both  traditional and modern approaches \\ to software updates in vehicles.\end{tabular} &
  \begin{tabular}[c]{@{}l@{}}The paper discusses dynamic \\ software updating in embedded \\ systems, emphasizing routine-based\\  and component-based updates.\end{tabular} \\ \hline
\begin{tabular}[c]{@{}l@{}}Technical Challenges \\ Addressed\end{tabular} &
  \begin{tabular}[c]{@{}l@{}}Implementing dynamic software \\ updates due to the complexity of \\ automotive E/E architecture, \\ security vulnerabilities, and the \\ need for robust and reliable systems.\end{tabular} &
  \begin{tabular}[c]{@{}l@{}}Discusses the challenges of OTA \\ updates, including energy consumption, \\ device memory usage, network \\ traffic overhead, and the balance \\ between performance gains and \\ update costs.\end{tabular} &
  \begin{tabular}[c]{@{}l@{}}Include implementing OTA updates \\ in automotive systems, integrating these\\  updates safely and efficiently, and \\ protecting Electronic Control Units\\ (ECUs) from external intrusions.\end{tabular} &
  \begin{tabular}[c]{@{}l@{}}Challenges in updating real-time\\  systems due to high operational risks \\ and costs, and the necessity of updates\\  to adapt to new communication \\ standards and address software \\ complexities \\ and potential bugs.\end{tabular} \\ \hline
\begin{tabular}[c]{@{}l@{}}Advantages and \\ Limitations\end{tabular} &
  \begin{tabular}[c]{@{}l@{}}Advantages: OTA updates reduce \\ downtime and improve efficiency. \\ Virtualization maximizes hardware \\ resource use. Container-based \\ architectures offer scalability. \\ Limitations: Security risks with OTA \\ updates, complexity in virtualization, \\ and implementation challenges.\end{tabular} &
  \begin{tabular}[c]{@{}l@{}}Advantages: OTA updates facilitate \\ remote modifications and enhance \\ device functionality. \\ Limitations: High energy consumption, \\ increased complexity in development and \\ deployment, resource-intensive update \\ steps, and significant overhead in device \\ memory and network.\end{tabular} &
  \begin{tabular}[c]{@{}l@{}}While OTA updates offer fast software \\ delivery and dynamic upgrading \\ capabilities, they also require \\ sophisticated management and face \\ challenges in development and \\ flexibility of software architectures.\end{tabular} &
  \begin{tabular}[c]{@{}l@{}}Advantages: Dynamic updating enhances\\  flexibility, and real-time operating \\ systems like Wind River VxWorks \\ indirectly support these updates. \\ Limitations: Challenges \\ in real-time systems due to deadlines \\ and resources; real-time OSes often \\ lack inherent support for updates, \\ and the Component Framework adds \\ design complexity.\end{tabular} \\ \hline
\begin{tabular}[c]{@{}l@{}}Innovations and \\ Solutions Proposed\end{tabular} &
  \begin{tabular}[c]{@{}l@{}}Container-based virtualization and \\ the integration of advanced \\ encryption methods. Suggests \\ enhancing E/E architectures and \\ exploring lightweight virtualization \\ solutions.\end{tabular} &
  \begin{tabular}[c]{@{}l@{}}Proposes sophisticated protocols and \\ techniques for OTA updates, \\ focusing on security, efficient \\ dissemination, and reliable \\ installation.\end{tabular} &
  \begin{tabular}[c]{@{}l@{}}New approaches for integrating IoT \\ technologies with Adaptive AUTOSAR, \\ focusing on safety, security, and efficient \\ cloud connectivity in the update process.\end{tabular} &
  \begin{tabular}[c]{@{}l@{}}Proposes operating system features \\ and a component framework approach \\ for dynamic updates in real-time systems.\\  Operating systems like Wind \\ River VxWorks offer features like remote \\ debugging, while the Component \\ Framework allows dynamic replacement\\  of software modules.\end{tabular} \\ \hline
\begin{tabular}[c]{@{}l@{}}Security and Safety\\  Considerations\end{tabular} &
  \begin{tabular}[c]{@{}l@{}}The need for secure software updates\\  to address vulnerabilities in \\ connected vehicles and the robustness \\ of container-based systems.\end{tabular} &
  \begin{tabular}[c]{@{}l@{}}Highlights the need for advanced security \\ measures to protect data integrity and \\ confidentiality during OTA updates.\end{tabular} &
  \begin{tabular}[c]{@{}l@{}}Emphasizes the importance of robust \\ security measures to protect ECUs and \\ ensure OTA updates do not compromise \\ vehicle safety.\end{tabular} &
  \begin{tabular}[c]{@{}l@{}}Need for secure update mechanisms in \\ real-time systems, though specific security \\ considerations are not deeply explored.\end{tabular} \\ \hline
\begin{tabular}[c]{@{}l@{}}Implementation \\ Complexity\end{tabular} &
  \begin{tabular}[c]{@{}l@{}}Complexity in integrating virtualization \\ technologies like containers and \\ unikernels into automotive software.\end{tabular} &
  \begin{tabular}[c]{@{}l@{}}Discusses the complexity involved in \\ managing OTA updates, including the\\ challenges in security measures and \\ efficient dissemination strategies.\end{tabular} &
  \begin{tabular}[c]{@{}l@{}}Acknowledges the complexity involved \\ in integrating OTA updates with \\ automotive systems, particularly in \\ terms of safety and security.\end{tabular} &
  \begin{tabular}[c]{@{}l@{}}Implementing updates in real-time systems, \\ particularly with the Component Framework \\ and ensuring compatibility with existing \\ operating systems.\end{tabular} \\ \hline
Performance Metrics &
  \begin{tabular}[c]{@{}l@{}}Focuses on reducing vehicle downtime,\\  improving hardware resource efficiency, \\ and enhancing operational scalability and \\ efficiency.\end{tabular} &
  \begin{tabular}[c]{@{}l@{}}Focuses on minimizing energy consumption,\\  optimizing device memory \\ usage, and managing network traffic \\ effectively during OTA updates.\end{tabular} &
  \begin{tabular}[c]{@{}l@{}}Focuses on the efficiency, speed of \\ software delivery, and the safety and \\ security of the update process in \\ automotive systems.\end{tabular} &
  \begin{tabular}[c]{@{}l@{}}Maintaining system performance during \\ updates, with a goal to overcome \\ current limitations that cap system \\ performance at a maximum frequency \\ of 300 Hz.\end{tabular} \\ \hline
Compliance with Standards &
  \begin{tabular}[c]{@{}l@{}}Discusses alignment with automotive \\ industry standards, particularly in the \\ context of connected and autonomous \\ vehicle systems.\end{tabular} &
  \begin{tabular}[c]{@{}l@{}}Considers the importance of aligning \\ OTA update processes with IoT standards \\ for security, efficiency, and reliability.\end{tabular} &
  \begin{tabular}[c]{@{}l@{}}Discusses the need to align software\\  updates with automotive industry standards, \\ especially concerning IoT integration.\end{tabular} &
  \begin{tabular}[c]{@{}l@{}}Aligning dynamic software updates with \\ industry standards in power and \\ automation sectors, particularly for \\ real-time embedded systems.\end{tabular} \\ \hline
\end{tabular}
}
\end{table*}

%% file: tables_tex/table2.tex
\setlength\extrarowheight{12pt}
\begin{table*}[ht!]
\centering
\resizebox{.95\textwidth}{!}{%
\begin{tabular}{|l|l|l|l|l|}
\hline
{\bf Comparison Matrix $\downarrow$} &
  {\bf Details from Mugarza et al. Paper $\downarrow$} &
  {\bf Details from Ghosal et al. Paper $\downarrow$} &
{\bf Details from Chi et al. Paper $\downarrow$} &
  {\bf Details from Kolomvatsos Paper $\downarrow$} \\ \hline
Application Domain &
  \begin{tabular}[c]{@{}l@{}}Focuses on IoT and IIoT systems, \\ with a special emphasis on smart \\ city applications and high \\ availability energy management \\ systems.\end{tabular} &
  \begin{tabular}[c]{@{}l@{}}The paper focuses on IoT environments, particularly \\ on secure Over-the-Air (OTA) update techniques for \\ connected vehicles.\end{tabular} &
  \begin{tabular}[c]{@{}l@{}}The paper discusses dynamic software\\  updating approaches in wireless sensor \\ networks (WSNs), particularly for IoT \\ applications.\end{tabular} &
  \begin{tabular}[c]{@{}l@{}}Focuses on wireless sensor networks (WSNs) \\ within IoT, especially on the dynamic and \\ decentralized nature of these environments.\end{tabular} \\ \hline
Update Methodology &
  \begin{tabular}[c]{@{}l@{}}Discusses Dynamic Software Updating \\ (DSU) techniques, particularly in \\ the context of continuous operation \\ without system shutdowns.\end{tabular} &
  \begin{tabular}[c]{@{}l@{}}Discusses secure OTA update techniques, \\ including scalable scheduling, \\ use of cryptographic hash functions, symmetric \\ and asymmetric key cryptography, and application \\ of Attribute-Based Encryption (ABE) \\ techniques.\end{tabular} &
  \begin{tabular}[c]{@{}l@{}}Focuses on dissemination-based approaches, \\ traffic reduction strategies, and the role of the \\ execution environment in DSU.\end{tabular} &
  \begin{tabular}[c]{@{}l@{}}Discusses a distributed approach to software \\ updating, emphasizing incremental updates, \\ autonomous update mechanisms, and distributed \\ management schemes.\end{tabular} \\ \hline
\begin{tabular}[c]{@{}l@{}}Technical Challenges \\ Addressed\end{tabular} &
  \begin{tabular}[c]{@{}l@{}}The challenges of implementing DSU in\\  high-availability applications, \\ emphasizing the need for updates during \\ runtime and the importance of \\ addressing security vulnerabilities and \\ privacy concerns.\end{tabular} &
  \begin{tabular}[c]{@{}l@{}}Highlights challenges in ensuring security during \\ OTA updates in connected vehicles, including issues \\ like authentication, confidentiality, integrity, and \\ data freshness.\end{tabular} &
  \begin{tabular}[c]{@{}l@{}}Addresses the challenges of efficiently\\  distributing update packets in \\ environments without pre-established \\ infrastructure, optimizing bandwidth and \\ energy through software patches, and \\ ensuring proper operation within existing \\ hardware and software infrastructure of sensor\\  networks.\end{tabular} &
  \begin{tabular}[c]{@{}l@{}}Addresses the challenges of updating software \\ in dynamic, decentralized IoT environments, \\ including issues with incremental updates, node \\ autonomy, and managing updates across \\ distributed systems.\end{tabular} \\ \hline
\begin{tabular}[c]{@{}l@{}}Advantages and \\ Limitations\end{tabular} &
  \begin{tabular}[c]{@{}l@{}}Advantages: Enables updates during runtime,\\  enhances system availability, and addresses \\ security and privacy. \\ Limitations: Implementation complexity, \\ particularly in integrating these systems and\\  managing component criticality levels.\end{tabular} &
  \begin{tabular}[c]{@{}l@{}}Advantages: Improved management of update \\ schedules, enhanced security through cryptographic \\ measures, and data integrity maintenance. \\ Limitations: Some methodologies lack comprehensive \\ scalable scheduling, challenges in source reliability \\ detection in blockchain-based solutions, and complexities\\  in achieving comprehensive encryption.\end{tabular} &
  \begin{tabular}[c]{@{}l@{}}Advantages: Efficient update dissemination,\\ energy and bandwidth conservation. \\ Limitations: Complexity in managing \\ dissemination and traffic reduction, and \\ ensuring compatibility with various hardware and \\ software configurations in sensor networks.\end{tabular} &
  \begin{tabular}[c]{@{}l@{}}Advantages: Minimizes disruption and resource \\ use, enhances node self-sufficiency, and improves\\  overall system resilience. \\ Limitations: Complexity in implementation, \\ potential inconsistencies in update \\ applications across the network.\end{tabular} \\ \hline
\begin{tabular}[c]{@{}l@{}}Innovations and \\ Solutions Proposed\end{tabular} &
  \begin{tabular}[c]{@{}l@{}}Proposes the Cetratus Framework for \\ Industrial Control Systems, \\ facilitating live updates aligned with \\ industrial standards, and employing an \\ indirection handling table for component \\ management.\end{tabular} &
  \begin{tabular}[c]{@{}l@{}}Proposes various methodologies like Secure Software \\ Repository Frameworks (e.g., Uptane), Blockchain-Based \\ Solutions (e.g., INFUSE), and STRIDE for enhancing \\ OTA update security.\end{tabular} &
  \begin{tabular}[c]{@{}l@{}}Proposes strategies like Directed Diffusion and \\ MOAP for dissemination, patch generation for traffic\\ reduction. Highlights virtual machines like Maté for\\ code efficiency and reprogramming simplification.\end{tabular} &
  \begin{tabular}[c]{@{}l@{}}Proposes a novel approach that includes a \\ performance-aware mechanism, load balancing \\ scheme, ensemble forecasting, and decision-making \\ using an Artificial Neural Network (ANN) for \\ optimizing software update \\ timings.\end{tabular} \\ \hline
\begin{tabular}[c]{@{}l@{}}Security and Safety \\ Considerations\end{tabular} &
  \begin{tabular}[c]{@{}l@{}}Emphasizes enhancing security and privacy in \\ smart city applications, including the integration \\ of a homomorphic encryption algorithm in a \\ data collector agent.\end{tabular} &
  \begin{tabular}[c]{@{}l@{}}Emphasizes the critical need for secure software \\ updates to mitigate risks in connected vehicles, \\ focusing on aspects like authentication, \\ confidentiality, integrity, and data freshness.\end{tabular} &
  \begin{tabular}[c]{@{}l@{}}While focusing on update efficiency and system\\ functionality, the paper does not deeply delve \\ into the specific security and safety aspects of \\ the DSU process in WSNs.\end{tabular} &
  \begin{tabular}[c]{@{}l@{}}While focusing on efficiency and autonomy, the\\ paper does not deeply delve into specific security \\ and safety considerations in the DSU \\ process.\end{tabular} \\ \hline
Implementation Complexity &
  \begin{tabular}[c]{@{}l@{}}Highlights the complexity involved in applying \\ DSU techniques in high-availability systems, \\ especially the integration of the Cetratus \\ Framework.\end{tabular} &
  \begin{tabular}[c]{@{}l@{}}Notes the complexity involved in implementing \\ secure OTA updates, especially when integrating \\ advanced cryptographic techniques and \\ ensuring end-to-end security.\end{tabular} &
  \begin{tabular}[c]{@{}l@{}}Discusses the complexities involved in \\ implementing DSU techniques in WSNs, especially\\ regarding managing network traffic and ensuring \\ compatibility with diverse sensor network \\ environments.\end{tabular} &
  \begin{tabular}[c]{@{}l@{}}Notes the complexity involved in implementing \\ a distributed, autonomous, and performance-aware \\ software update system in WSNs.\end{tabular} \\ \hline
Performance Metrics &
  \begin{tabular}[c]{@{}l@{}}Focuses on ensuring continuous operation and \\ addressing resource management challenges, \\ especially in terms of energy consumption and \\ operational efficiency.\end{tabular} &
  \begin{tabular}[c]{@{}l@{}}Focuses on the efficiency and security of the \\ OTA update process, including aspects like update \\ delivery speed, system reliability, and \\ data protection.\end{tabular} &
  \begin{tabular}[c]{@{}l@{}}Emphasizes the importance of efficient software \\ dissemination, reduced bandwidth and energy usage, \\ and maintaining the operational functionality of \\ sensor networks during updates.\end{tabular} &
  \begin{tabular}[c]{@{}l@{}}Emphasizes efficient, timely, and autonomous \\ software updates in IoT environments, with a \\ focus on optimizing performance metrics like update\\ timing and network load.\end{tabular} \\ \hline
Compliance with Standards &
  \begin{tabular}[c]{@{}l@{}}Discusses the need to align DSU methods with \\ industrial standards, ensuring compatibility and \\ reliability in smart city and industrial IoT \\ applications.\end{tabular} &
  \begin{tabular}[c]{@{}l@{}}Discusses the importance of aligning OTA update \\ methods with security standards in the automotive \\ industry, ensuring the safety and \\ reliability of connected vehicles.\end{tabular} &
  \begin{tabular}[c]{@{}l@{}}This paper does not explicitly focus on compliance \\ with specific standards but addresses the general \\ requirements and best practices for \\ DSU in wireless sensor networks within IoT contexts.\end{tabular} &
  \begin{tabular}[c]{@{}l@{}}This paper does not explicitly focus on compliance \\ with specific standards but addresses the general \\ requirements for efficient software \\ updates in distributed IoT environments.\end{tabular} \\ \hline
\end{tabular}
}
\caption{Comparative Analysis}
\label{table1}
\end{table*}